\documentclass[11pt,twoside]{article}
\usepackage{asp2010}
\usepackage{graphicx}

\resetcounters

\newcommand{\sla}{\raisebox{-0.10em}{$\stackrel{<}{{\mbox{\tiny $\sim$}}}$}}

\begin{document}

\title{Metal-Abundance Peculiarities in the sdOB Star EC\,11481$-$2303:\\ A Progress Report}
\author{Ellen Ringat and Thomas Rauch}{\small \affil{Institute for Astronomy and Astrophysics, Kepler Center for Astro and Particle Physics, Eberhard Karls University, Sand 1, 72076 T\"ubingen, Germany}}

\begin{abstract}
Spectral analyses of extreme horizontal branch (EHB) stars indicate a trend to supersolar heavy-metal abundances that is increasing with effective temperature. The sdOB EC 11481$-$2303 is such a star, located at the very hot end of the EHB. In contrast to other sdOB stars, it is exhibiting an extraordinary flat UV continuum. This may be the result of extreme metal-line blanketing. We present preliminary results of a spectral analysis by means of  static as well as stratified NLTE model atmospheres.
\end{abstract}

\section{Introduction}
Subdwarf B (sdB) stars form a homogeneous class of extreme horizontal-branch (EHB) stars (20\,000\,K $\sla \  T_\mathrm{eff}\  \sla\  40\,000\,$K, 5 $\sla\  \log g \ \sla\  6$) that have evolved directly from the red giant branch. They are core helium-burning stars with hydrogen envelopes thin enough to prevent hydrogen burning. Their spectra show broad Balmer lines up to a principle quantum number of $n \approx 12$ and mostly no helium lines. sdOB stars show similar spectra but additionally a weak He\,\textsc{ii} $\lambda\,4686\,\AA\,$ line. ´They are found at higher $T_\mathrm{eff}$ ($>40\,000$\,K).  

In the last few years, many abundance analyses of subdwarf stars were performed. E.g.\@ \citet{Edelmann2} analyzed a sample of 146 sdB and sdOB stars and found constant abundance values for similar $T_\mathrm{eff}$ and $\log g$ but a general trend of an increasing helium abundance with rising $T_\mathrm{eff}$. Three stars showed extremely overabundant iron-group elements (up to 32\,000\,$\times$ solar). 
\citet{Geier} performed an abundance analysis of 139 sdB stars and detected a trend of increasing heavy-element abundances (generally supersolar) with $T_\mathrm{eff}$ whereas iron itself is not affected. This analysis showed that the super metal-rich stars of \citet{Edelmann2} are not that peculiar considering the supersolarity of heavy-metal abundances. In fact such values can be explained with atomic diffusion (equilibrium between gravitational and radiative forces), that becomes more effective at higher $T_\mathrm{eff}$ and confirms the trend found by \cite{Geier}.
 
There are still some features that cannot be explained. If pulsations are considered, iron is believed to play a role in the driving mechanism. Therefore precise determination of iron-group abundances are important to understand the difference between pulsating and non-pulsating stars and to explain the pulsation processes.

One very peculiar star is EC\,11481$-$2303, a relatively hot sdOB star ($T_\mathrm{eff}=55\,000\,$K) that shows extreme overabundances of iron and nickel and exhibits a peculiar, flat UV continuum that is not explained. Its atmosphere seems to be dominated by iron-group elements and therefore, it is an ideal candidate to test static and stratified models as well as stellar evolutionary theory. Spectral analysis of such a metal-rich star is still very difficult, mainly because of the lack of reliable atomic data. We summarize previous analyses in Sect.\,\ref{discovery}. Our own analysis procedure is described in Sect.\,\ref{method} considering static as well as stratified model atmospheres, and the interstellar line-absorption spectrum. Ways to access the \emph{TMAP} model atmosphere code and its atomic data are shown in Sect.\,\ref{GAVO}. We conclude with our preliminary results in Sect.\,\ref{results}.

\section{Discovery and Previous Analyses}
\label{discovery}

Discovered in the Edinburgh-Cape Blue Objects survey, EC\,11481$-$2303 is a relatively bright source with $m_\mathrm{V}=11.76$ \citep{Kilkenny} and was classified as a DA white dwarf with a companion 6$ \farcs$6 away. 

\citet{Stys} performed a first spectral analysis with LTE models considering hydrogen and helium only. They used optical SAAO (Aug 15, 1995, 1.9m telescope, 3350-5450\,$\AA\,$, resolution 3.5\,$\AA\,$) and IUE spectra (Jul 14, 1993, 1150-1978\,$\AA\,$, resolution 7\,$\AA\,$ for SWP48111 and 0.1\,$\AA\,$ for SWP48112). From the optical spectra they determined $T_\mathrm{eff}=41\,790$\,K, $\log g=5.84$ and  He/H=0.014 (number fraction), but were not able to fit the UV continuum with these models. From the UV spectra, they determined $\log n_{\mathrm{H}\textsc{i}}=20.5$ and concluded that EC\,11481$-$2303 must have the same distance (500\,pc) and reddening (0.06) as the nearby (distance $4 \fdg 4$) star HD\,104377 for which the same $\log n_{\mathrm{H}\textsc{i}}$ and similar line-of-sight elements and their velocities were determined by \citet{VanSteenberg}. Considering this distance, \citet{Stys} determined an absolute visual magnitude of $M_\mathrm{V}=3.2$ for EC\,11481$-$2303 which is typical for a subdwarf star. Therefore EC\,11481$-$2303 was re-classified as subdwarf star.

In order to explain the peculiar flat UV continuum, \citet{Stys} considered the analysis of the DAB white dwarf GD\,323 performed by \citet{Koester}. GD\,323 shows a similar flat continuum-flux shape. They considered models with homogeneously mixed composition, chemically stratified constitution, a He spot at the surface, and a composite of a DA and a DB white dwarf. 
Only stratified models were able to fit the flux shape, but those models did not match the Balmer lines. Therefore the flux shape was still unexplained and \citet{Stys} concluded that EC\,11481$-$2303 is worth a follow-up study.

Within the framework of the ESO Supernovae Type Ia Progenitor Survey (SPY) project \citep{Napiwotzki}, a sdO sample was observed with the UVES spectrograph at the VLT and analyzed by \citet{Stroeer}. EC\,11481$-$2303 was excluded from this analysis among other things because of its UV flux shape. However they suggested that EC\,11481$-$2303 has a higher $T_\mathrm{eff}$ and lower helium abundance than determined by \citet{Stys}.
This SPY spectrum was used also by \citet{Jordan} to investigate magnetic fields in DA white dwarfs, but no hints of a magnetic field were found in the observations of EC\,11481$-$2303.

\citet{Rauch} used this spectrum together with the previously mentioned IUE and additional FUSE observations. With models calculated using the T\"ubingen NLTE Model-Atmosphere Package (\emph{TMAP}, \citealt{Werner}, \citealt{Rauch-Deetjen}) containing the elements H, He, C, N, O, and (Ca-Ni), they determined $T_\mathrm{eff}=55\,000\,$K, $\log g=5.8$, He/H=0.0025 by number, and upper limits [C]\,=\,$-$0.875, [N]\,=\,$-$0.875, and [O]\,=\,$-$0.875 from the optical spectrum ([X] denotes log [abundance / solar abundance] of element X).
To match the UV flux shape, they concluded that the abundances for the iron-group elements must be at least ten times solar and that nickel seems to be even more overabundant. 
They suggested that diffusion calculations should be performed in a next step and the iron-group elements should be considered individually (they used a generic model atom). The reddening should be redetermined, and interstellar lines should be modeled. 
All this was picked up for this analysis. The exact procedure is described in the next section.

\section{Analysis procedure}
\label{method}

In the following, we introduce the model-atmosphere code for static model atmospheres, the statistical treatment of the iron-group elements, and the program for stratified atmospheres. Then we describe the line-fitting procedure and the program for the interstellar lines.

\subsection{\emph{TMAP} - The Model-Atmosphere Code}

\emph{TMAP}\footnote{http://astro.uni-tuebingen.de/\raisebox{.2em}{\scriptsize $\sim$}TMAP/TMAP.html} has been developed since the 1980s and is well tested in many spectral analyses (e.g.\@ \citealt{Wassermann}). It assumes hydrostatic and radiative equilibrium and can consider plane-parallel or spherical geometry. \emph{TMAP} provides reliable results for temperatures between $20\,000\,$K$ \ \sla\ \ T_\mathrm{eff} \ \sla\  200\,000$\,K and $4 \ \sla\  \log g \ \sla\  9$ and up to 1500 NLTE levels can be included. An input for this code is model atoms from the T\"ubingen Model Atom Database (\emph{TMAD}\footnote{http://astro.uni-tuebingen.de/\raisebox{.2em}{\scriptsize $\sim $}TMAD/TMAD.html}). This database was built up together with the model atmosphere code, therefore its format is \emph{TMAP} compliant. It is continuously updated. Presently it contains data for the elements H, He, C, N, O, F, Ne, Na, Mg, Si, S, Ar, and Ca. Data for other elements are in progress. Model atoms for model-atmosphere calculations as well as for SED calculations (that include fine-structure splitting) can be downloaded.

\subsection{\emph{IrOnIc} - Statistical Treatment For Iron-Group Elements}

Iron-group elements (Ca - Ni) consist of hundreds of millions of lines due to their electron configuration (partly filled 3d and 4s shells). Therefore a statistical treatment is necessary to consider all lines but to get a reasonable number of NLTE levels. This treatment is implemented in the Iron Opacity and Interface code (\emph{IrOnIc}, \citealt{Rauch-Deetjen}) that divides the excitation-energy range of an ion into several bands and combines all atomic levels within one band to superlevels. The line-absorption cross-sections are sampled to superlines, ranging from one band to another or within one band. With this treatment all levels are considered, but the number of NLTE levels is strongly reduced.

The program uses POS and LIN lines from Kurucz' line lists\footnote{http://kurucz.harvard.edu/atoms.html} as input. POS lines contain only observed (good) wavelengths, LIN lines include theoretical as well as observed wavelengths and there are therefore about 300 times more of them.

\begin{table}
\begin{center}
\caption{Apparent magnitudes of EC\,11481$-$2303 from SIMBAD.}
\label{mag}
\begin{tabular}{ccccc}
\hline
B&V&J&H&K\\
11.49&11.76&12.512&12.669&12.765\\
\hline
\end{tabular}
\end{center}
\end{table}

\articlefigure[angle=-90, width=\textwidth]{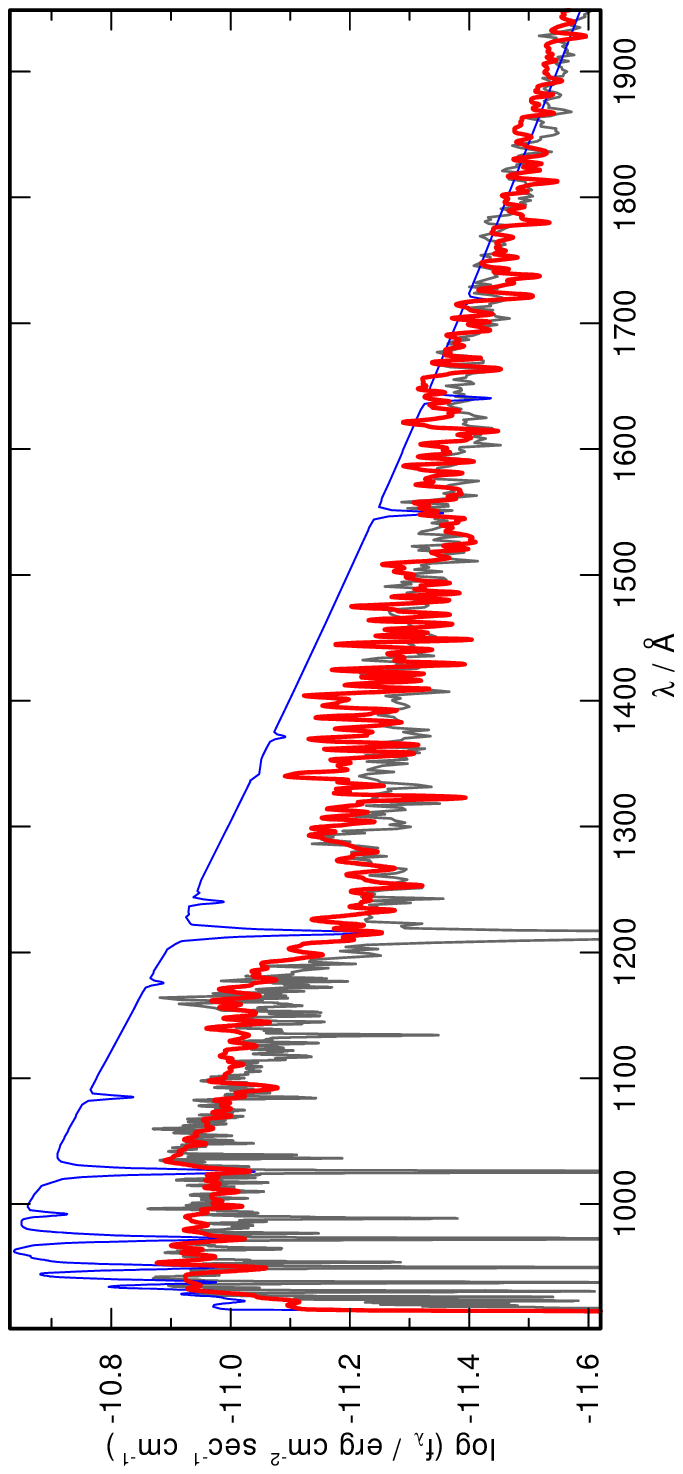}{flux}{The UV spectrum of EC\,11481$-$2303 compared to a model containing HHeCNO (blue, thin line) and hydrogen and iron-group elements with 10\,$\times$ solar iron and 1000\,$\times$ solar nickel value (red, thick line). Gaussians of 1\AA\, and 3\,$\AA\,$ (FWHM) are applied to the FUSE observation and synthetic spectrum for clarity. The synthetic spectra are normalized to the reddest magnitude (Table\,\ref{mag}). A reddening of $E_\mathrm{B-V}=0.03$ (0.01 for the HHeCNO model) is applied.}

At first, we calculated a HHeCNO model atmosphere with the values determined by \citet{Rauch}. To re-determine the reddening, we used the FUSE and IUE spectra as well as optical and infrared magnitudes from SIMBAD\footnote{http://simbad.u-strasbg.fr} (Table\,\ref{mag}). We normalized the synthetic flux to the $m_\mathrm{K}$ magnitude and applied the law of \citet{Fitzpatrick}. A value of $E_\mathrm{B-V}=0.06$ as determined by \citet{Stys} is too high. With our models $E_\mathrm{B-V}=0.02\pm0.01$ fits best.

For this analysis, we started with a model atom including hydrogen and all iron-group elements individually. We adopted $T_\mathrm{eff}$ and $\log g$ from \citet{Rauch} and increased the abundances of single iron-group elements up to 1000\,$\times$ solar. With these models, we compared our synthetic spectra with the observations (FUSE: B0540901 from May 2001 and the previously mentioned IUE and UVES observations) and found that only the elements Ni and Fe are able to decrease the flux shape notable in the relevant regions (Fig.\,\ref{flux}). We tested combinations of these elements with different supersolar abundances. The flux is reproduced best by a model with 1000\,$\times$ solar nickel abundance and 10-100\,$\times$ solar iron abundance. The increase of the other iron-group elements to supersolar values to this model did not reduce the flux significantly.  

\subsection{\emph{NGRT} - The Interplay of Radiative Levitation and Gravitational Settling}

The atmospheres of sdOB stars are dominated by diffusion. We used the Next Generation Radiative Transport (\emph{NGRT}, \citealt{Dreizler}) code that considers this effect. The theory was originally developed by \citet{Chayer} and modified by \citet{Dreizler}. \emph{NGRT} uses the \emph{TMAP} model as input. In an iteration cycle, the radiative acceleration is calculated in a fixed atmosphere structure, then the radiation field and the atmospheric structure are computed again, etc.\\\\

\articlefigure[angle=-90,width=\textwidth]{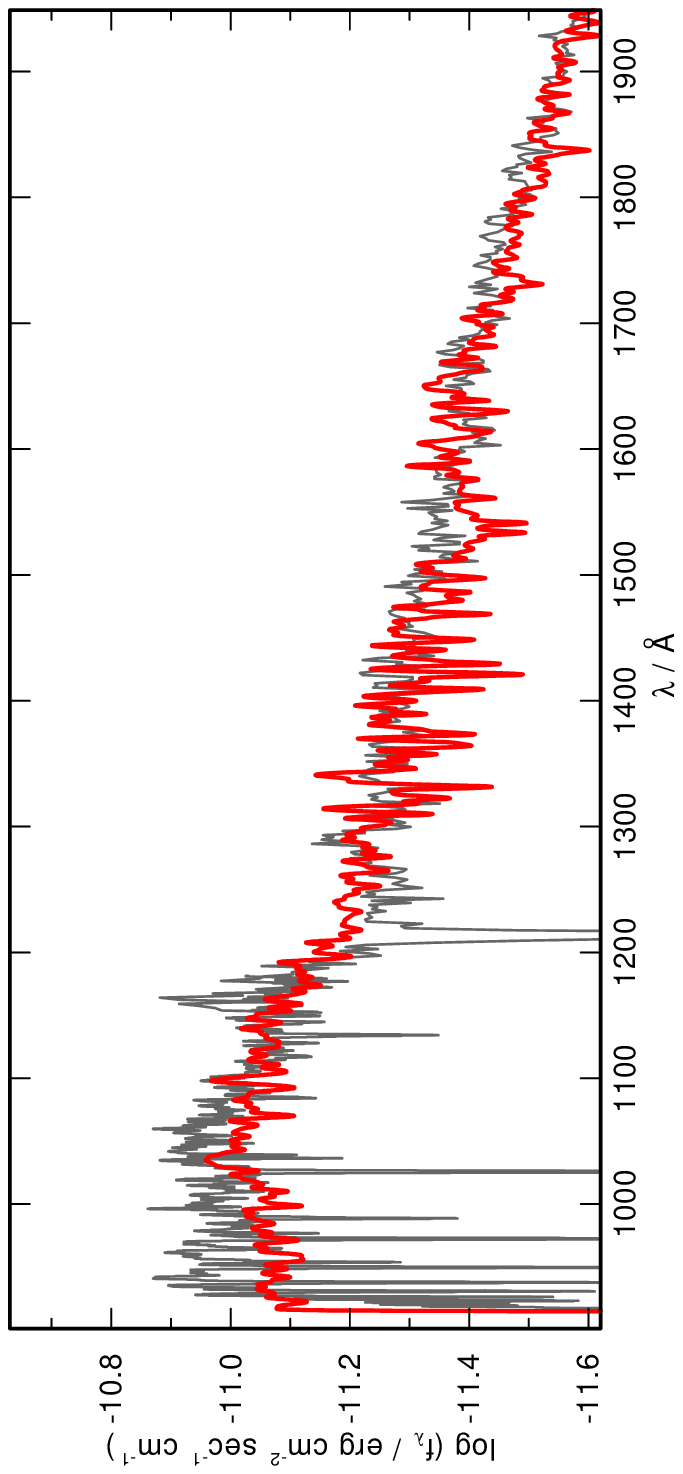}{diff}{Same as Fig.\,\ref{flux}. The model considers diffusion and $E_\mathrm{B-V}=0.025$.}

With the model containing hydrogen and the iron-group elements with solar abundances as start model, the \emph{NGRT} calculation provided the abundance distribution shown in Fig.\,\ref{diff}. This model reproduces the observation in many parts of the wavelength range. It also confirms the results determined with the \emph{TMAP} model, both yield (extreme) supersolar abundances for iron and nickel (Fig.\,\ref{abund}) and show a flat UV continuum.

\articlefigure[width=\textwidth]{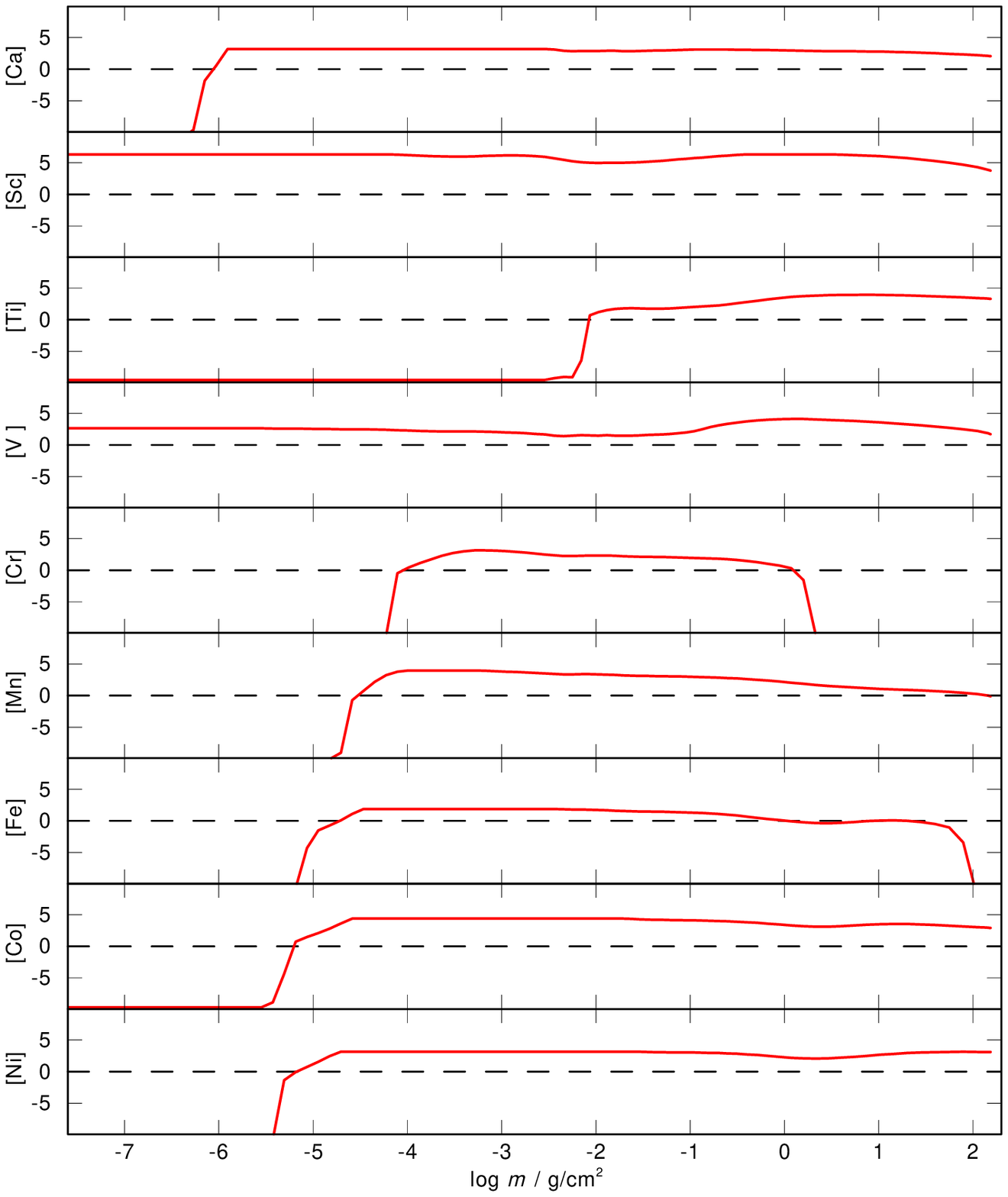}{abund}{Depth-dependent abundances of the single iron-group elements in our \emph{NGRT} diffusion model. [X] denotes log [abundance / solar abundance] of element X. The solar abundance is displayed by the dashed line.}

\subsection{Detailed Modeling of the Line Profiles}
The FUSE and IUE observations show many deep and broad absorption features that seem to stem from blended finer lines. With a pure HHeCNO model, only a few lines can be identified. Most lines seem to originate from iron-group elements. 

In order to model the line profiles, we compared the best-fit static and stratified models with the FUSE and IUE observations. Surprisingly, models and observations seem to have nothing in common although some individual features match the position of the observed lines considering the \emph{TMAP} model (Fig.\,\ref{FUSE}). The same is valid for individual regions of the spectrum. E.g.\@ from $996-1004\,\AA\,$ the model can reproduce the observation while nearby regions of the spectrum do not match at all (Fig.\,\ref{FUSE}).

\articlefigure[angle=-90, width=\textwidth]{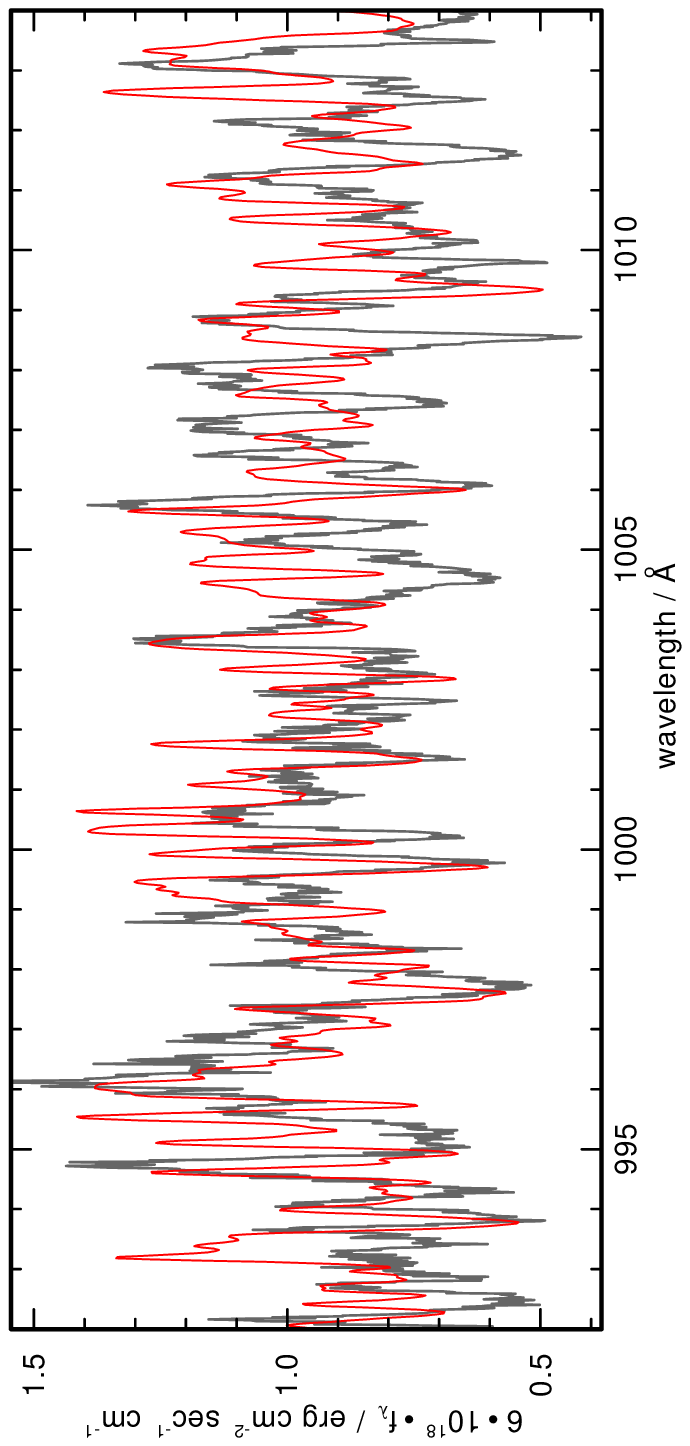}{FUSE}{Section of the FUSE observation compared with the best-fit \emph{TMAP} model. A Gaussian of 0.06\,\AA\, (FWHM), a radial velocity of $-17.5\,\mathrm{km}/\,\mathrm{sec}$, and a rotational velocity of $30\,{\mathrm{km}}/\,{\mathrm{sec}}$ are applied to the synthetic spectrum. A neutral hydrogen column density of 3\,$ \times 10^{20}$/ cm$^2$ and a reddening of $E_{B-V}=0.02$ are considered.} 

The theoretical lines are much narrower than observed. This could be a hint for stellar rotation. The rotational broadening smears out narrow lines and they appear as one broad line. In the analysis of \citet{Rauch}, the model shows a slightly higher emission reversal of the H$\alpha$ line than the optical observation. With a rotational velocity of $v_{rot}=30\,{\mathrm{km}}/\,{\mathrm{sec}}$ applied to their model (Fig.\,\ref{rot}), this emission feature matches the size of the observed line. With this $v_{rot}$, more features in the FUSE observation can be reproduced.
\citet{Rauch} included iron-group elements with a generic model atom into their calculations. In the fit to the Balmer lines, their model shows the so-called Balmer-line problem \citet{Napi-Rauch}. Thus, the heavy-metal abundances are still too low or not enough metal opacity is considered.

\articlefigure[angle=-90, width=\textwidth]{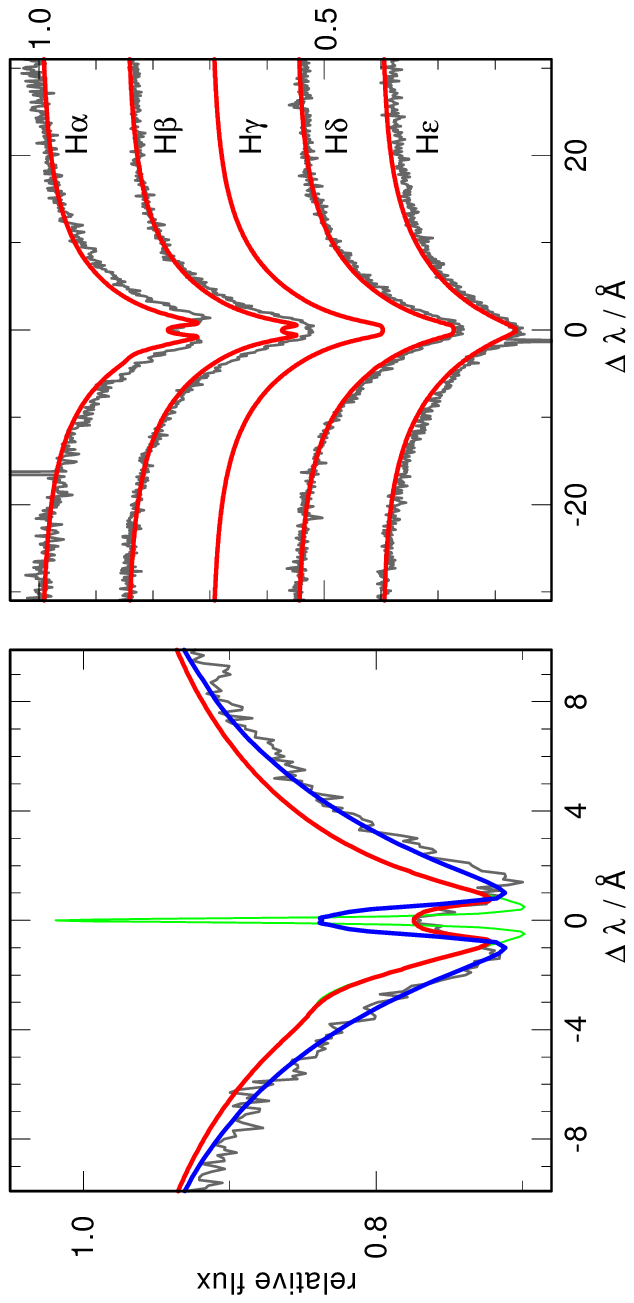}{rot}{The theoretical Balmer-line profiles with $v_{rot}=0$ and 30\,${\mathrm{km}}/\,{\mathrm{sec}}$ (green, thin and red, thick, respectively) compared with the observation. Additionally our best-fit model with 10\,$\times$ solar Fe and 1000\,$\times$ solar Ni abundance is displayed with $v_{rot}=30\,{\mathrm{km}}/\,{\mathrm{sec}}$ (blue).}

We calculated models with individually increased abundances of the iron-group elements. The comparison with the observation shows that no line can be unambiguously identified. The lines mainly stem from Fe and Ni, but cannot be reproduced perfectly. The reason is most likely that the Kurucz' LIN lines bear some uncertainties. We use Kurucz' line lists as input and only a few wavelengths are observationally confirmed (POS) for the high ionization stages that we have to consider in the case of EC\,11481$-$2303. About 300 times more lines are only theoretically calculated (LIN) with uncertainties of $\approx$10\% in wavelength and $\approx$15\% in f-values. All LIN lines have to be included in the calculations because otherwise a huge amount of opacity is missing. But this could be the reason that the lines do not fit perfectly or seem to be shifted a bit.

\subsection{\emph{OWENS} - Modeling Interstellar Absorption Lines}

To model the interstellar line-absorption spectrum, we use the program \emph{OWENS}. It can model different interstellar clouds with different temperatures, radial and turbulent velocities, and column densities. The resulting ISM spectrum is combined with our \emph{TMAP} spectrum and some of the strong lines can be identified as interstellar (Fig.\,\ref{ism}). But still the majority of the lines is of photospheric origin.

\articlefigure[angle=-90, width=\textwidth]{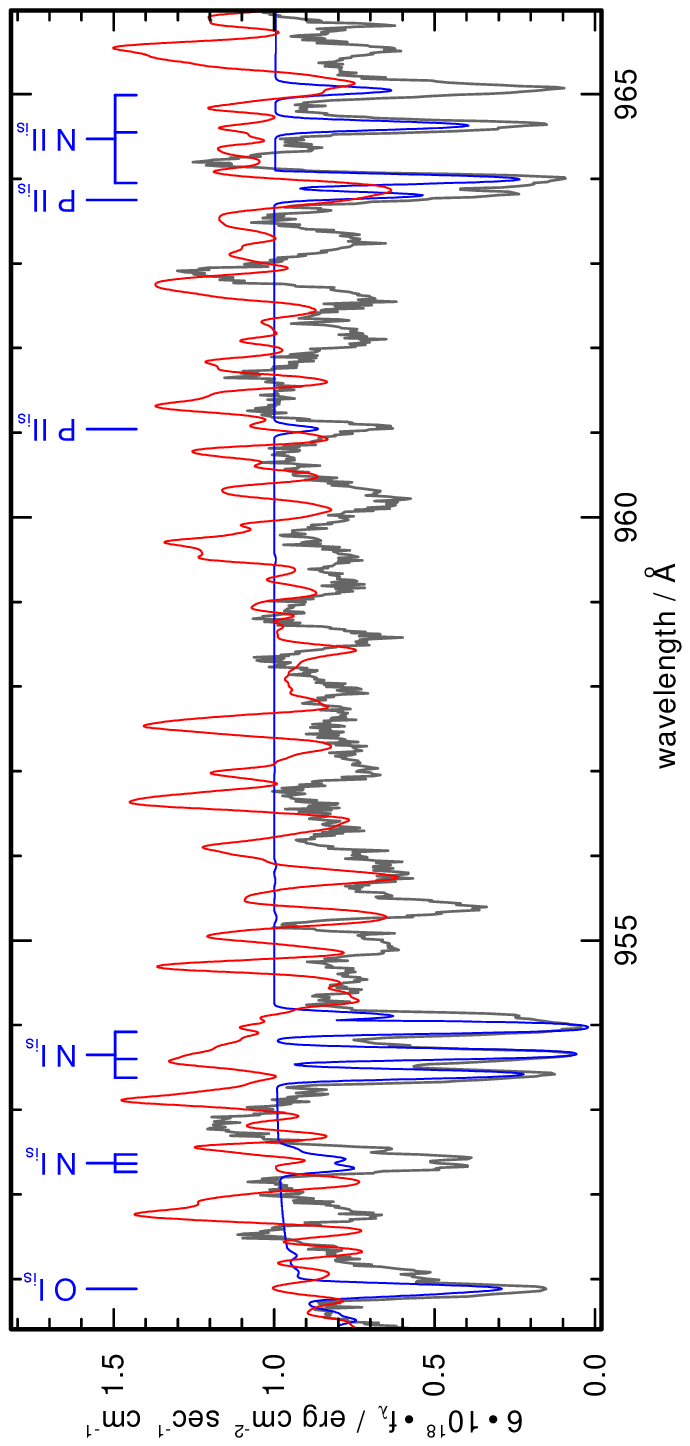}{ism}{Section of the FUSE observation compared with the best-fit \emph{TMAP} model (red) and the normalized \emph{OWENS} ISM model (blue). The \emph{TMAP} model has the same applications as in Fig.\,\ref{FUSE}.} 

\section{\emph{TheoSSA} - Virtual Observatory}
\label{GAVO}

The German Astrophysical Virtual Observatory (\emph{GAVO}\footnote{http://www.g-vo.org/}) aims to make astronomical data of all kinds accessible worldwide. It is part of the International Virtual Observatory Alliance (\emph{IVOA}\footnote{http://www.ivoa.net/}) and is a contact point for everybody needing assistance with publishing his or her own data or using tools or services. 

Within the framework of a \emph{GAVO} project, \emph{TMAP} was made accessible:\\
1) Already calculated spectral energy distributions (SEDs) can be downloaded via the service \emph{TheoSSA} (http://dc.g-vo.org/theossa).\\
2) Individual models can be calculated with the web-interface of \emph{TMAP} with the service \emph{TMAW} (http://astro.uni-tuebingen.de/\raisebox{.2em}{\scriptsize $\sim$}TMAW/TMAW.shtml).\\
3) Ready-to-use model atoms can be downloaded and used for every code or tailored individually from the T\"ubingen Model Atom Database\\ (\emph{TMAD}, http://astro.uni-tuebingen.de/\raisebox{.2em}{\scriptsize $\sim $}TMAD/TMAD.html).

We will include the final SED calculated for this analysis into \emph{TheoSSA}.

\newpage
\section{Preliminary Results and Conclusion}
\label{results}

We calculated static and stratified model atmospheres to reproduce the peculiar, flat UV continuum flux shape of EC\,11481$-$2303. The elements H, Ca, Sc, V, Cr, Mn, Fe, Co, and Ni are considered and we increased the abundances of the single elements individually and for some combinations in the static calculations. The best-fit model contains about 1000\,$\times$ solar nickel and about 10-100\,$\times$ solar iron (Table\,\ref{abundtab}). The other iron-group elements have almost no influence on the overall flux shape.

\begin{table}
\caption{Parameters of EC\,11481$-$2303. $T_\mathrm{eff}$, $\log g$ and the abundances for H-O are adopted from \citet{Rauch}. IG denotes Ca\,-\,Mn and Co.}
\label{abundtab}
\small{
\begin{tabular}{llllllllll}
\hline
$T_\mathrm{eff}$/\,K& $\log g$& [H] & [He] & [C] &[N]&[O]&[IG]&[Fe]&[Ni]\\
55\,000&5.8&0.127&$-$1.405&$-$0.875&$-$0.875&$-$0.875&1-100&10-100&1000\\
\hline
\end{tabular}
}
\end{table}

The stratified model yielded abundances for the iron-group element that are all supersolar. This confirms the abundance results from the static model. Both the stratified and the static models can reproduce most parts of the observed UV flux shape.

If the absorption lines of the IUE and FUSE observations are considered, both models are unable to reproduce these lines. From the fit to the H$\alpha$ line, a rotational velocity of 30\,${\mathrm{km}}/\,{\mathrm{sec}}$ can be determined. Applying this to the models in the UV range improves the fit and some parts of the observation are matched. 

The interstellar line absorption spectrum was modeled with the program \emph{OWENS} and some of the strongest lines are of interstellar origin. 
But still, the atmosphere seems to be dominated by iron-group elements. To model those lines, atomic data must be reliable. At the moment no such data are available for temperatures in the range of hot sdOB stars ($T_\mathrm{eff}>40\,000$\,K). Most lines are theoretically calculated and have uncertainties in the wavelength range as well as in the f-values. We suggest that these uncertainties are the reason for the discrepancy between the models and the observation.

\acknowledgements  ER is supported by the Federal Ministry of  Education and Research (BMBF) under grant 05A11VTB. TR is supported by the German Aerospace Center (DLR) under grant 05\,OR\,0806. This work was done using the profile fitting procedure OWENS.f developed by M. Lemoine and the \emph{FUSE} French Team. We gratefully thank the bwGRiD project\footnote{bwGRiD (http://www.bw-grid.de), member of the German D-Grid initiative,
      funded by the Ministry for Education and Research (Bundesministerium fuer
      Bildung und Forschung) and the Ministry for Science, Research and Arts
      Baden-Wuerttemberg (Ministerium fuer Wissenschaft, Forschung und Kunst
      Baden-Wuerttemberg).} for the computational resources.

\end{document}